\documentclass[journal]{IEEEtran}
\usepackage{cite}
\usepackage{amsmath,amssymb,amsfonts}
\usepackage{graphicx}
\usepackage{xcolor}
\usepackage{siunitx}
\usepackage{booktabs}

\definecolor{notepink}{RGB}{231,41,138}

\begin{document}

\title{High Characteristic Voltage YBCO Josephson Junctions Written With a Focused Helium Ion Beam}

\author{Rochelle~Y.~Qu,~\IEEEmembership{Member,~IEEE,} Aravind~R.~Sreeja,
Adhilsha~Parachikunnumal, Alexander~Brooks,~\IEEEmembership{Member,~IEEE,} Nicole~Y.~Chin,~\IEEEmembership{Member,~IEEE,} Jay~C.~LeFebvre,~\IEEEmembership{Member,~IEEE,} Roger~K.~Lake,~\IEEEmembership{Fellow,~IEEE,} and Shane~A.~Cybart,~\IEEEmembership{Senior~Member,~IEEE}
\thanks{This paper is based upon work supported by the NNSA
Minority Serving Institutions Partnership Program under Award DE-NA0004106 and AFOSR grant FA9550-23-1-0369.}
\thanks{The authors are with the Department of
Electrical and Computer Engineering, University of California, Riverside, CA 92521 USA (email: \mbox{cybart@ucr.edu}).}}

\markboth{IEEE TRANSACTIONS ON APPLIED SUPERCONDUCTIVITY}%
{Qu \MakeLowercase{\textit{et al.}}: High Characteristic Voltage YBCO Josephson Junctions Written With a Focused Helium Ion Beam}

\maketitle

\begin{abstract}
We report in-plane Josephson junctions written directly into YBa$_2$Cu$_3$O$_{7-\delta}$ (YBCO) thin films with a focused helium ion beam (He FIB). The characteristic voltage $I_{\mathrm{C}}R$ of these junctions, where $I_{\mathrm{C}}$ is the Josephson critical current and $R$ is the junction resistance, sets the operating frequency and the sensitivity of the circuits built from them. A single line irradiation with a dose of \SI{500}{ions/nm} across a \SI{4}{\micro\metre} wide electrode produced a junction with $I_{\mathrm{C}}R$ of 929 \si{\micro\volt} at \SI{1.6}{K}, which is `higher than the 200 to 400 \si{\micro\volt} typical of junctions fabricated previously in our laboratory. Temperature dependent resistance, current-voltage characteristics, magnetic field modulation of the critical current, and Shapiro steps under 20.5~GHz microwave irradiation show Josephson behavior with an excess current below 5 percent of the critical current. The critical current falls steeply with temperature near the junction transition and saturates at low temperature. We attribute this behavior to a barrier at the metal insulator transition, where hopping conduction through localized states is active at finite temperature and freezes out on cooling. We describe the changes in surface preparation, sample selection, beam focusing, and sample storage that accompanied the improvement, although the present data do not separate their individual contributions. Larger characteristic voltages at these dimensions are an encouraging result for rapid single flux quantum and adiabatic quantum flux parametron logic and for superconducting quantum interference devices operating at elevated temperature.
\end{abstract}

\IEEEpeerreviewmaketitle


\section{Introduction}
\label{sec:intro}
\IEEEPARstart{S}{ince} the discovery of high-transition temperature (high-$T_{\mathrm{C}}$) superconductivity in YBa$_2$Cu$_3$O$_{7-\delta}$ (YBCO), researchers have worked to develop a Josephson junction technology for superconducting electronics, detectors, and cryogenic circuits. The transition temperature of YBCO is above the boiling point of liquid nitrogen, which substantially reduces the cooling requirements compared with low-$T_{\mathrm{C}}$ superconductors such as Nb.\cite{Hilgenkamp2002,Likharev1979} However, a higher operating temperature also means higher thermal noise. Junctions with a large characteristic voltage are therefore needed to keep bit error rates low in rapid single flux quantum (RSFQ) and adiabatic quantum flux parametron (AQFP) logic, and to preserve the sensitivity of superconducting quantum interference devices (SQUIDs).\cite{Koelle.1999,Cybart.2008,Cho.2015} In-plane high-$T_{\mathrm{C}}$ superconductor (HTS) Josephson junctions remain difficult to make because the transport is anisotropic, the material is sensitive to oxygen loss, and the superconducting properties depend strongly on disorder on nanometer length scales.\cite{Hilgenkamp2002,Hammerl2000,Qu.2025,LeFebvre.2023}

A key figure of merit for a Josephson junction is the characteristic voltage
\begin{equation}
V_{\mathrm{C}} = I_{\mathrm{C}}R,
\end{equation}
where $I_{\mathrm{C}}$ is the Josephson critical current and $R$ is the junction resistance. The characteristic voltage sets the Josephson frequency scale and therefore the maximum operating speed of superconducting circuits. It also governs the microwave response, dynamic range, and sensitivity of Josephson devices.\cite{Barone1982,Likharev1986} Raising $I_{\mathrm{C}}R$ while keeping the junctions uniform is therefore a central goal for high performance superconducting electronics.

The value quoted for $I_{\mathrm{C}}R$ depends on how $I_{\mathrm{C}}$ and $R$ are defined. This matters for HTS junctions, whose barriers span the metal insulator transition. A critical current read directly from the current-voltage ($\mathit{IV}$) characteristic can include excess current and thermal rounding, and a resistance taken at an arbitrary bias point can differ from the barrier resistance. In this work we define $I_{\mathrm{C}}$ for the temperature series as the current at which the voltage across the junction exceeds a criterion of 5 \si{\micro\volt} in the $\mathit{IV}$ characteristic measured inside a magnetic shield. We define $R$ as the slope of the $\mathit{IV}$ characteristic in the high-bias regime where it is linear, and the extrapolation of that line to zero voltage gives the excess current $I_{\mathrm{exc}}$. The amplitude of the critical current modulation in an applied magnetic field, which isolates the phase coherent Josephson current, is used as an independent check of $I_{\mathrm{C}}$ and as a bound on the excess current.

Ion irradiation has been used to make HTS Josephson junctions since White et al. showed in 1988 that ion damage destroys superconducting phase coherence in YBCO.\cite{White.1988} More recently, focused helium ion beam irradiation has become an established method for writing Josephson junctions directly into YBCO thin films.\cite{Cybart.2015,Cybart.2014,Cho.2018} The key to this technique is that YBCO is very sensitive to disorder from irradiation. The ion beam suppresses superconductivity locally by introducing nanoscale disorder, and the barrier is set by the dose alone, with no grain boundary or etched gap. At low dose the irradiated region is a transparent metallic weak link with superconductor-normal metal-superconductor (SNS) behavior. At high dose it becomes a resistive tunnel barrier with superconductor-insulator-superconductor (SIS) behavior. This tunability makes He FIB junctions useful for studying how $I_{\mathrm{C}}$, $R$, excess current, and magnetic field modulation change with barrier transparency.\cite{LeFebvre.2023,Propper.2026,Schmid.2025}

In this paper, we report in-plane He FIB Josephson junctions in YBCO with a higher $I_{\mathrm{C}}R$ than we have achieved before. Section~\ref{sec:methods} describes the film, the patterning and junction writing, and the measurement configuration, and notes the steps that differed from our earlier process. Section~\ref{sec:results} presents the temperature dependence of the resistance, the $\mathit{IV}$ characteristics, the magnetic field modulation of the critical current, and the microwave response. The last three measurements together separate the Josephson current from any excess current. The present data do not identify which process change is responsible for the improvement, and we state this where the comparison is made.

\section{Experimental Methods}
\label{sec:methods}

\subsection{Film Growth and Patterning}

\begin{figure}[!t]
\centering
\includegraphics[width=\linewidth]{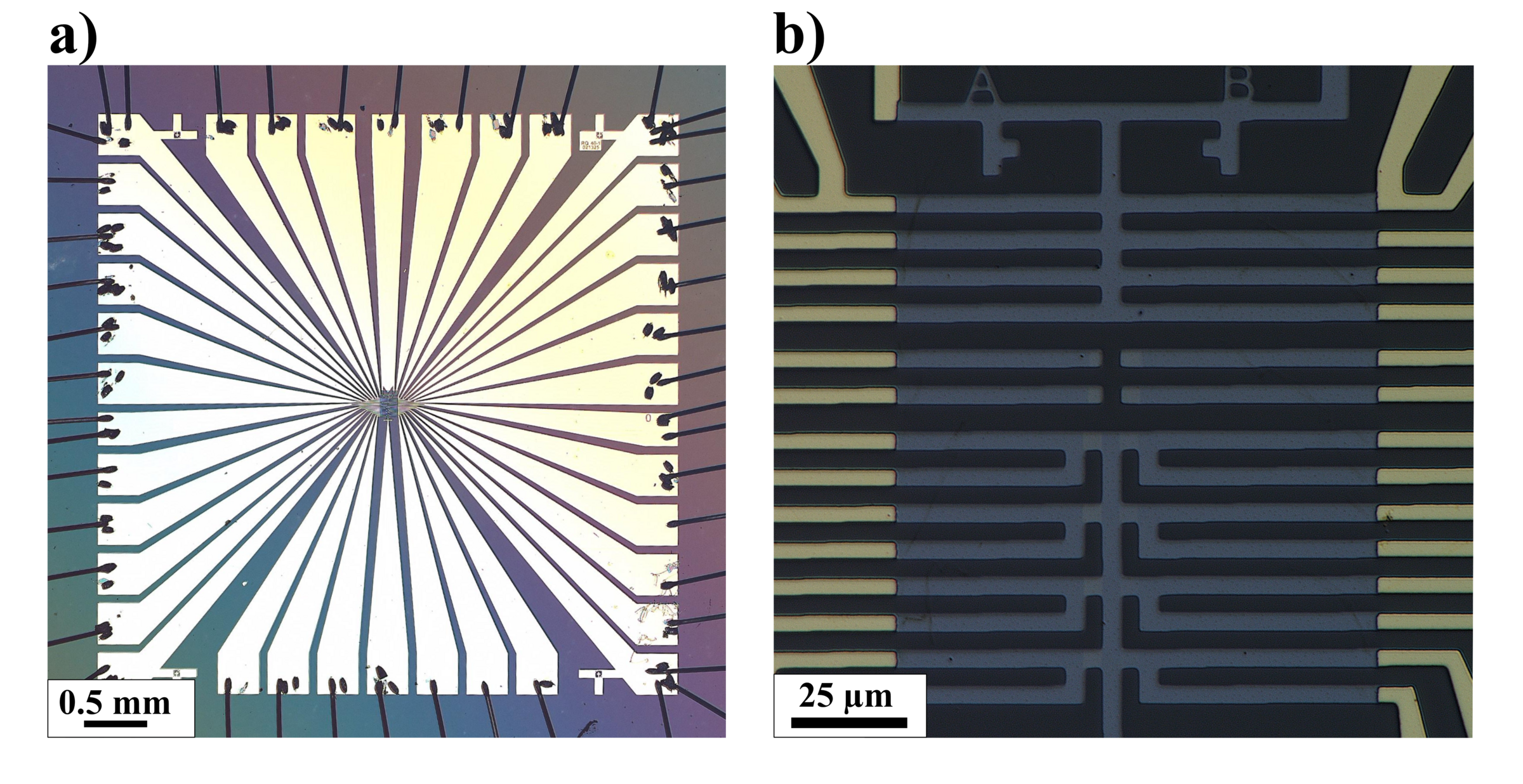}
\caption{(a) Optical micrograph of the \qtyproduct{5 x 5}{mm} chip patterned with gold (Au) contacts. Aluminum (Al) wire bonds connect the contact pads to the sample holder. (b) Magnified optical micrograph of the YBCO device layout for in-plane Josephson junctions, showing the Au contacts, the exposed YBCO electrodes, and the sapphire substrate. The layout includes both line irradiation and area irradiation test structures.}
\label{fig:design}
\end{figure}

The samples were \SI{40}{\nano\metre} thick epitaxial YBCO films grown by reactive coevaporation on CeO$_2$ buffered single crystal sapphire substrates, with an Au layer deposited in situ.\cite{Utz.1997,Wang.2019} The Au cap serves as a low resistance contact and protects the YBCO surface during patterning. The wafers were diced along the $a$ and $b$ directions of the YBCO crystal, so that the electrodes, and the junctions written across them, are oriented along either the $a$ or the $b$ axis. Before lithography the samples were cleaned by ultrasonic agitation in acetone and then in isopropanol, and each sample was inspected under an optical microscope. Only the cleanest samples, with no particulates or surface defects in the device region, were used for junction fabrication.

Figure~\ref{fig:design}(a) shows the resulting \qtyproduct{5 x 5}{mm} chip. The electrode geometry is designed for four probe measurements, and the chip is wire bonded to the sample holder. The magnified view in Fig.~\ref{fig:design}(b) shows the \SI{4}{\micro\metre} wide YBCO electrodes across which the junctions are written. The layer stack is sapphire, CeO$_2$, YBCO, and Au.

Patterning used two lithography steps with AZ Mir 701 photoresist exposed by a Microtech laser writer. In the first step the electrode pattern was transferred by argon ion milling, which removes the Au and YBCO in the exposed regions down to the sapphire and isolates the devices electrically. As in our usual process, the ion beam was shuttered during the etch to let the sample cool. Here it was shuttered for a longer time to limit plasma hardening of the photoresist, so that it stripped cleanly in acetone afterward. In the second step the Au was removed from the junction regions while the contact pads were kept. We chose a dilute potassium iodide (KI) solution for this step because it etches the Au without etching the YBCO. After this etch the samples were cleaned again to remove all resist and etch residue from the exposed YBCO, and they were inspected optically before irradiation. In this manner the superconducting surface was exposed only immediately before junction writing.

\subsection{Junction Writing}

Josephson junctions were written by rastering a He FIB across the exposed YBCO electrodes in a Zeiss Orion NanoFab helium ion microscope.\cite{10.1116/6.0004020} The tungsten source had been replaced within a month of writing these junctions, because a fresh source gives a brighter and more stable beam that focuses to a smaller spot. The accelerating voltage was \SI{35}{kV}, and the beam current was \SI{0.3\pm0.05}{pA} with a \SI{5}{\micro\metre} aperture and a spot control setting of 4, which is a machine setting of the microscope column and not a measure of the physical beam size. Before writing, the beam was focused and corrected for astigmatism. This optimization was performed within the write field, which is limited to \qtyproduct{100 x 100}{\micro\metre} to reduce beam deflection. The astigmatism was checked by burning a spot for \SI{1}{s} and imaging it with a \qtyproduct{100 x 100}{nm} field of view at a \SI{10}{\micro\second} dwell time. The burned spot was approximately \SI{5}{nm} in diameter, and a circular image of it confirms that the astigmatism has been corrected. The beam diameter itself was measured by scanning the beam across a sharp edge and was approximately \SI{0.5}{nm}. This step sets the junction geometry, because the barrier width and uniformity follow from the beam profile. Each junction was written as a single line across the \SI{4}{\micro\metre} wide electrode with a line dose of \SI{500}{ions/nm}. Patterning was controlled by Fibics Inc. NanoPatterning and Visualization Engine (NPVE) software with a spot dwell time of \SI{1}{\micro\second}, a point-to-point spacing of \SI{0.5}{nm}, and a nominal spot size of \SI{2}{nm}. The nominal spot size is the value NPVE uses to compute the dose per point and not the measured beam diameter. These settings result in a rastered exposure along the line. The irradiation introduces nanoscale disorder that suppresses superconductivity locally, and the line dose sets the barrier transparency. After writing, the samples were kept under vacuum in the microscope until they were measured, and they were stored in a dry box between measurements.

Four steps differed from our earlier fabrication runs. First, the exposed YBCO was cleaned after the second lithography step. Second, only the cleanest samples were used. Third, the beam was focused and corrected for astigmatism in every write field, and the spot was verified by imaging a burned spot and by scanning across a sharp edge, with the NPVE dwell time, point spacing, and spot size held fixed. Fourth, the samples were stored with desiccant inside a sealed container.\cite{9347702} The present data do not separate their individual contributions to the junction properties.

\subsection{Electrical Measurements}

Electrical measurements were made in a four probe configuration in a Quantum Design Physical Property Measurement System (PPMS) between \SI{1.6}{K} and room temperature. The temperature dependence of the resistance and the $\mathit{IV}$ characteristics were measured with the sample inside a mu-metal shield. The shield was removed for the magnetic field modulation and microwave measurements, so those measurements were made in the Earth's field. Furthermore, the superconducting magnet of the PPMS can carry a remanent field after being energized, so the field at the sample is offset from the nominal applied field. The magnetic field was applied perpendicular to the film. Critical currents were extracted from the $\mathit{IV}$ characteristics using a voltage criterion of 5 \si{\micro\volt}. Microwaves at \SI{20.5}{GHz} were coupled to the sample through an antenna line.

\section{Results and Discussion}
\label{sec:results}

\subsection{Temperature Dependence of the Resistance}

\begin{figure}[!t]
\centering
\includegraphics[width=0.43\textwidth]{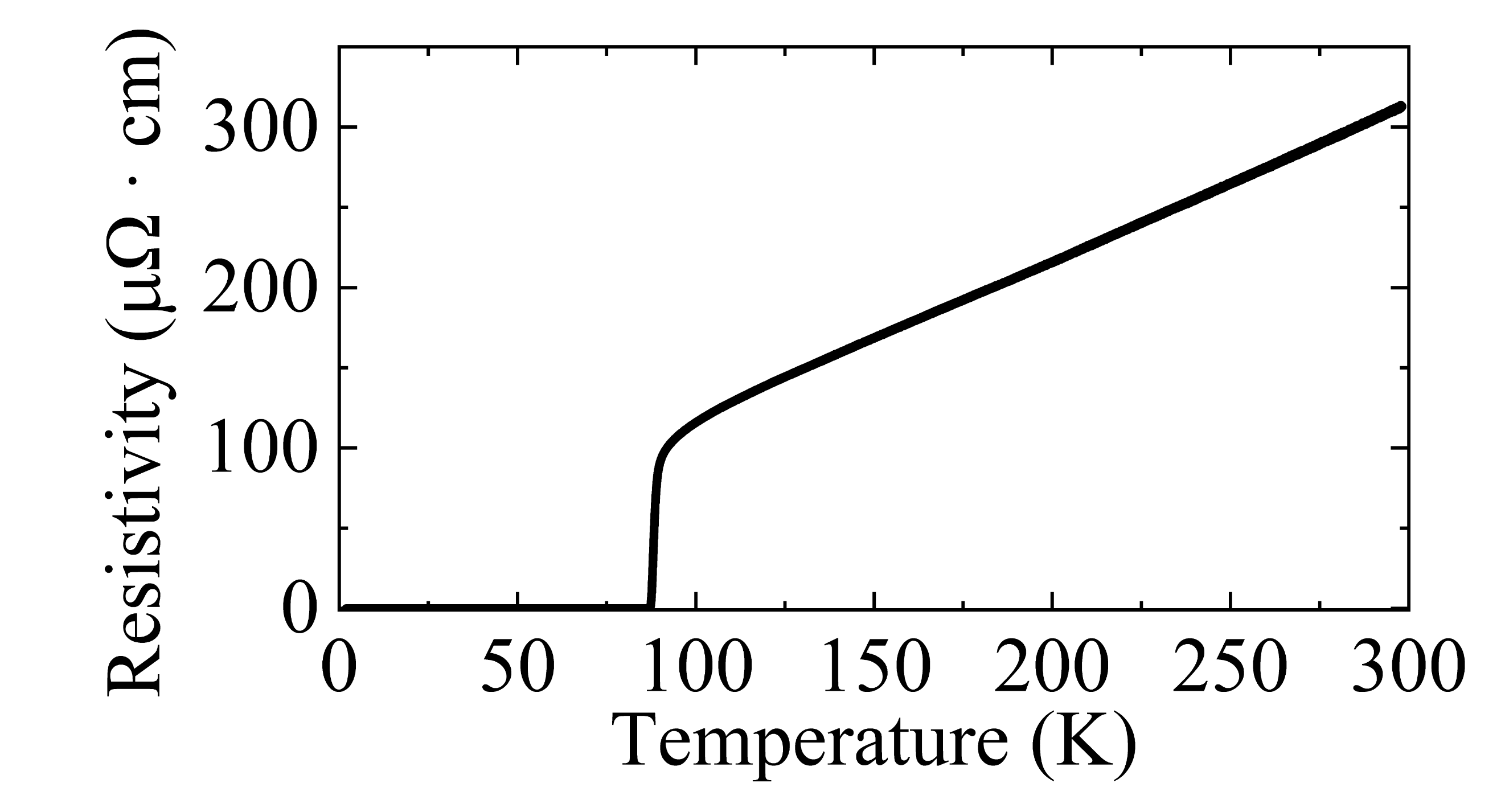}
\caption{Resistivity versus temperature of the unirradiated YBCO film up to \SI{300}{K}, measured with a current of \SI{400}{nA}, showing the superconducting transition near \SI{87}{K}.}
\label{fig:rt film}
\end{figure}

\begin{figure}[!t]
\centering
\includegraphics[width=0.45\textwidth]{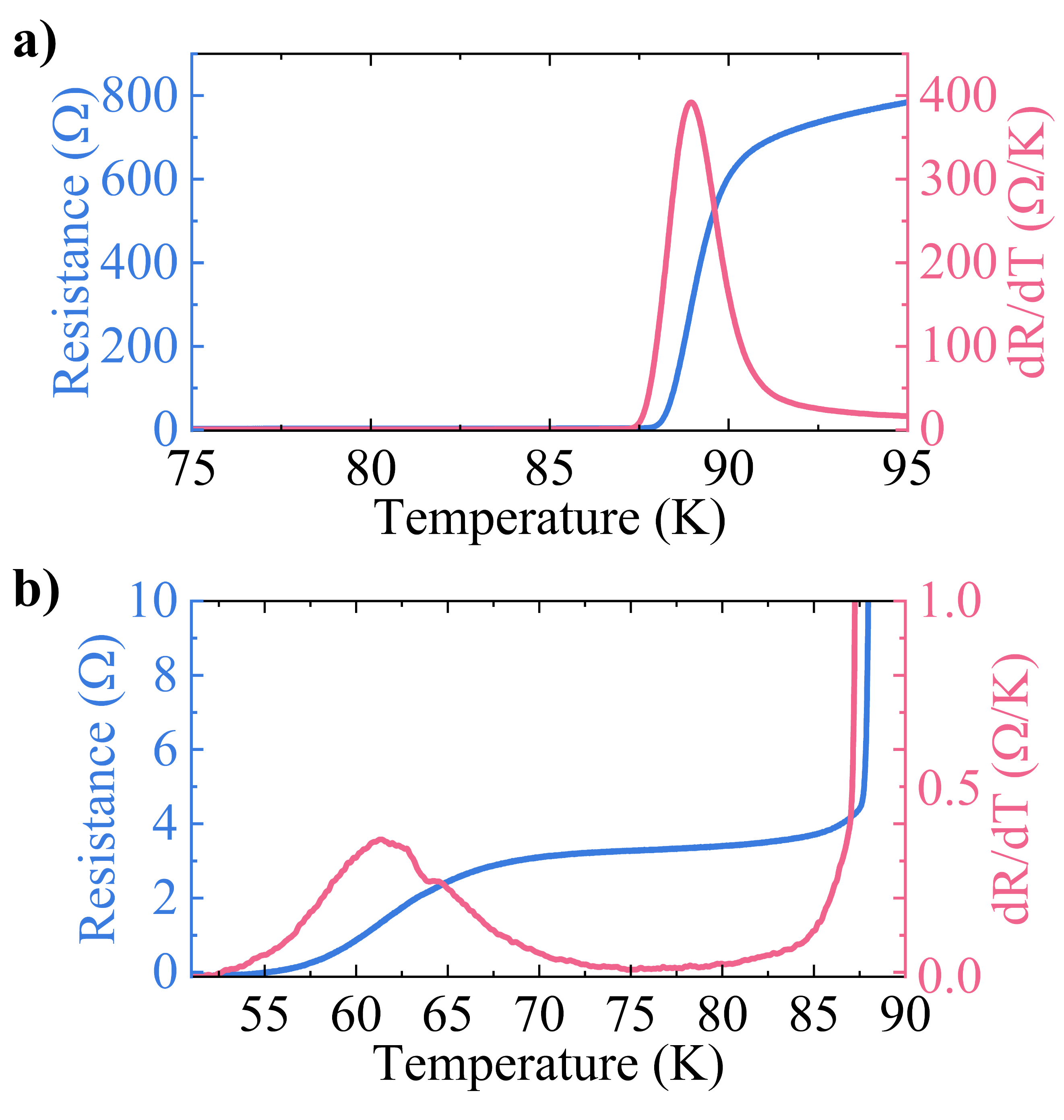}
\caption{Resistance $R$ (blue, left axis) and $dR/dT$ (red, right axis) versus temperature for a Josephson junction irradiated with a He dose of \SI{500}{ions/nm}. (a) Transition of the unirradiated YBCO electrodes near \SI{87}{K}. The width of the $dR/dT$ peak is a measure of the homogeneity of $T_\mathrm{C}$ across the electrodes. (b) Below the electrode transition the resistance settles onto a plateau of a few ohms set by the irradiated barrier and remains nearly constant with temperature. The resistance drops to zero near \SI{52}{K} once the Josephson critical current exceeds the \SI{400}{nA} measurement current.}
\label{fig:rt JJ}
\end{figure}

Resistivity versus temperature shows the quality of the unirradiated film and how far the irradiation suppresses superconductivity. Figure~\ref{fig:rt film} shows the resistivity of an unirradiated YBCO electrode measured with a current of \SI{400}{nA}. At high temperature the film is metallic with a nearly linear temperature dependence, and a sharp superconducting transition occurs at $T_\mathrm{C} = \SI{87\pm2}{K}$. Throughout this work we define each transition temperature as the temperature of the peak in $dR/dT$ versus $T$, and the quoted uncertainty is the full width at half maximum (FWHM) of that peak. The residual resistivity ratio $\rho(\SI{300}{K})/\rho(\SI{100}{K})$ is 2.62, which indicates good metallic transport above the transition.

Figure~\ref{fig:rt JJ} shows the resistance of an electrode containing an irradiated barrier, together with its temperature derivative $dR/dT$. Two features are visible. The first, in Fig.~\ref{fig:rt JJ}(a), is the transition of the unirradiated electrodes at \SI{87\pm2}{K}. Below this transition, in Fig.~\ref{fig:rt JJ}(b), the resistance does not vanish but settles onto a plateau of approximately \SI{3}{\ohm}, which is the resistance of the irradiated barrier in series with the superconducting electrodes. The plateau is nearly independent of temperature between roughly \SI{65}{K} and \SI{85}{K}. A metallic weak link would instead show a resistance that falls as the irradiated material is cooled. A flat plateau of a few ohms, with a weak upturn at low temperature, places the barrier on the insulating side of the metal insulator transition. We return to this point in Section~\ref{sec:iv}. The second feature, at \SI{52\pm1}{K} by the same peak and FWHM definition, is where the resistance drops to zero and $dR/dT$ returns to zero. Below this temperature the Josephson critical current exceeds the \SI{400}{nA} measurement current and the junction is in its zero voltage state. A measurable supercurrent therefore appears about \SI{35}{K} below the electrode transition. This reflects the reduction in superconducting order produced by the irradiation, whereas the electrodes themselves are unaffected.

\subsection{Current-Voltage Characteristics}
\label{sec:iv}

\begin{figure}[!t]
\centering
\includegraphics[width=0.44\textwidth]{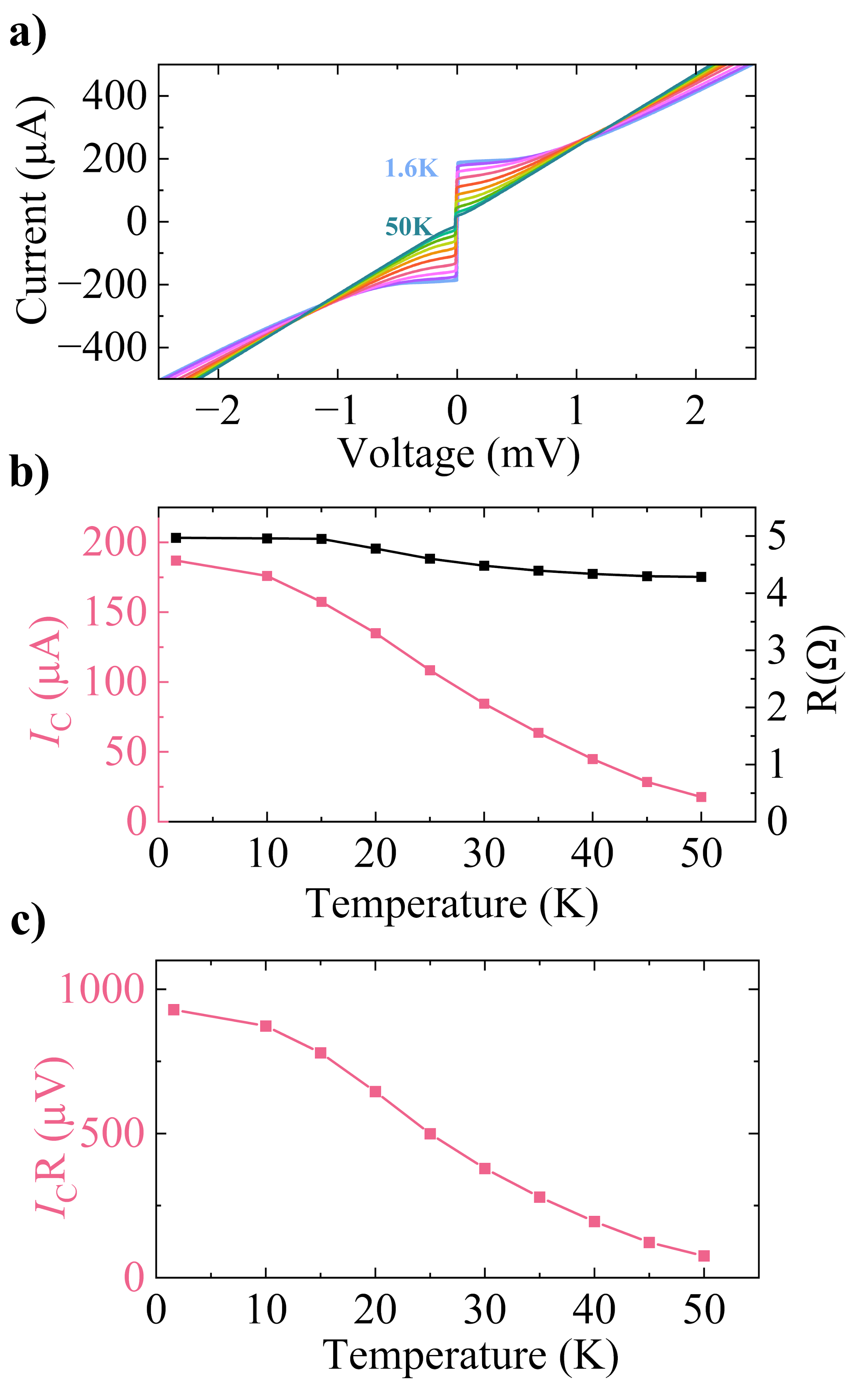}
\caption{(a) Current-voltage characteristics of the \SI{500}{ions/nm} Josephson junction measured at 1.6, 10, 15, 20, 25, 30, 35, 40, 45, and \SI{50}{K} inside a mu-metal shield. (b) Temperature dependence of $I_{\mathrm{C}}$ (red circles, left axis) and $R$ (black squares, right axis) extracted from the curves in (a). (c) Temperature dependence of the product $I_{\mathrm{C}}R$.}
\label{fig:IVT}
\end{figure}

Figure~\ref{fig:IVT}(a) shows the $\mathit{IV}$ characteristics of the junction measured inside the mu-metal shield between \SI{1.6}{K} and \SI{50}{K}. The curves have a pronounced zero voltage state and a sharp transition to the resistive branch. As the temperature increases, $I_{\mathrm{C}}$ decreases steadily while $R$ stays nearly constant, as summarized in Fig.~\ref{fig:IVT}(b). Here $R$ is the slope of a linear fit to the high-bias branch over the bias window from 1 to 2 \si{\milli\volt} at each temperature. The decrease of $I_{\mathrm{C}}$ with temperature follows from the thermal weakening of the order parameter and increased phase fluctuations.

The high-bias branches at all temperatures extrapolate to $I_{\mathrm{exc}} = $ 5 \si{\micro\ampere} at zero voltage, which corresponds to $I_{\mathrm{exc}}/I_{\mathrm{C}} = $ 0.03 at \SI{1.6}{K}. The excess current is therefore negligible, and the $\mathit{IV}$ characteristic is that of a tunnel-like barrier. Figure~\ref{fig:IVT}(c) shows the product $I_{\mathrm{C}}R$. It reaches 929 \si{\micro\volt} at \SI{1.6}{K} and remains measurable over the whole temperature range. These values are higher than those of the in-plane He FIB junctions previously measured in our laboratory, for which typical values of $I_{\mathrm{C}}R$ ranged from about 200 to 400 \si{\micro\volt} between 1.6 \si{K} and 50 \si{K}.

The temperature dependence of $I_{\mathrm{C}}$ in Fig.~\ref{fig:IVT}(b) has two regimes. Between about \SI{20}{K} and \SI{50}{K}, $I_{\mathrm{C}}$ falls nearly linearly with temperature, and a linear fit over this range extrapolates to zero near \SI{53}{K}, in good agreement with the \SI{52\pm1}{K} at which the resistance vanishes in Fig.~\ref{fig:rt JJ}(b). For a tunnel junction, Ambegaokar and Baratoff give
\begin{equation}
I_{\mathrm{C}}R = \frac{\pi\Delta(T)}{2e}\tanh\frac{\Delta(T)}{2k_BT},
\label{eq:AB}
\end{equation}
with $\Delta(T)$ the Bardeen-Cooper-Schrieffer (BCS) gap.\cite{Ambegaokar.1963} This form is flat over most of the temperature range and drops only close to the temperature where the gap closes. The measured falloff is far steeper than Eq.~(\ref{eq:AB}) for any choice of that temperature, and it resembles instead the proximity behavior of an SNS weak link. Below about \SI{15}{K}, $I_{\mathrm{C}}$ flattens and saturates, as Eq.~(\ref{eq:AB}) does. Since $R$ is nearly constant, $I_{\mathrm{C}}R$ in Fig.~\ref{fig:IVT}(c) has the same shape. Thermal noise rounds the apparent $I_{\mathrm{C}}R$ only where the Ambegaokar-Halperin parameter $\gamma = \hbar I_{\mathrm{C}}/ek_BT$ is small.\cite{Ambegaokar.1969} Here $\gamma$ is 20 at \SI{50}{K} and 90 at \SI{35}{K}, so thermal rounding is small over the range where $I_{\mathrm{C}}$ falls steeply in Fig.~\ref{fig:IVT}(b).

We attribute the two regimes to the barrier itself. The few ohm plateau in Fig.~\ref{fig:rt JJ}(b) places the irradiated YBCO at the metal insulator transition. In two dimensions this transition occurs near \SI{10}{k\ohm} per square.\cite{White.1988,Cybart.2014} The barrier is about \SI{1}{nm} long along the current and \SI{4}{\micro\metre} wide, or $2.5\times10^{-4}$ squares, which puts the transition at a few ohms. That is where these junctions sit. Material at this point is insulating only at zero temperature. At finite temperature it conducts by variable range hopping through localized states in the barrier.\cite{Mott.1969} At higher temperature this hopping channel is open, so the barrier behaves as a poor metal, the junction is a proximity coupled weak link, and $I_{\mathrm{C}}$ falls steeply with temperature. As the temperature drops, the hopping freezes out. Direct tunneling is then the only channel left, the junction becomes a tunnel junction, and $I_{\mathrm{C}}$ saturates. The weak increase of $R$ on cooling in Fig.~\ref{fig:IVT}(b), from about \SI{4.3}{\ohm} at \SI{50}{K} to \SI{5}{\ohm} at \SI{1.6}{K}, is the expected signature of the hopping channel freezing out. The negligible excess current at \SI{1.6}{K} is consistent with tunneling dominating there. The hopping channel also shunts the junction internally, which is why the $\mathit{IV}$ characteristics are not hysteretic. A barrier damaged far enough to be a band gap insulator would have a resistance of the order of \SI{10}{k\ohm} at these dimensions, and thermal fluctuations would then suppress the Josephson current. A barrier at the transition, with a resistance of a few ohms, is what gives both a measurable $I_{\mathrm{C}}$ and a large $I_{\mathrm{C}}R$.

\subsection{Magnetic Field Modulation and Excess Current}

\begin{figure}[!t]
\centering
\includegraphics[width=0.46\textwidth]{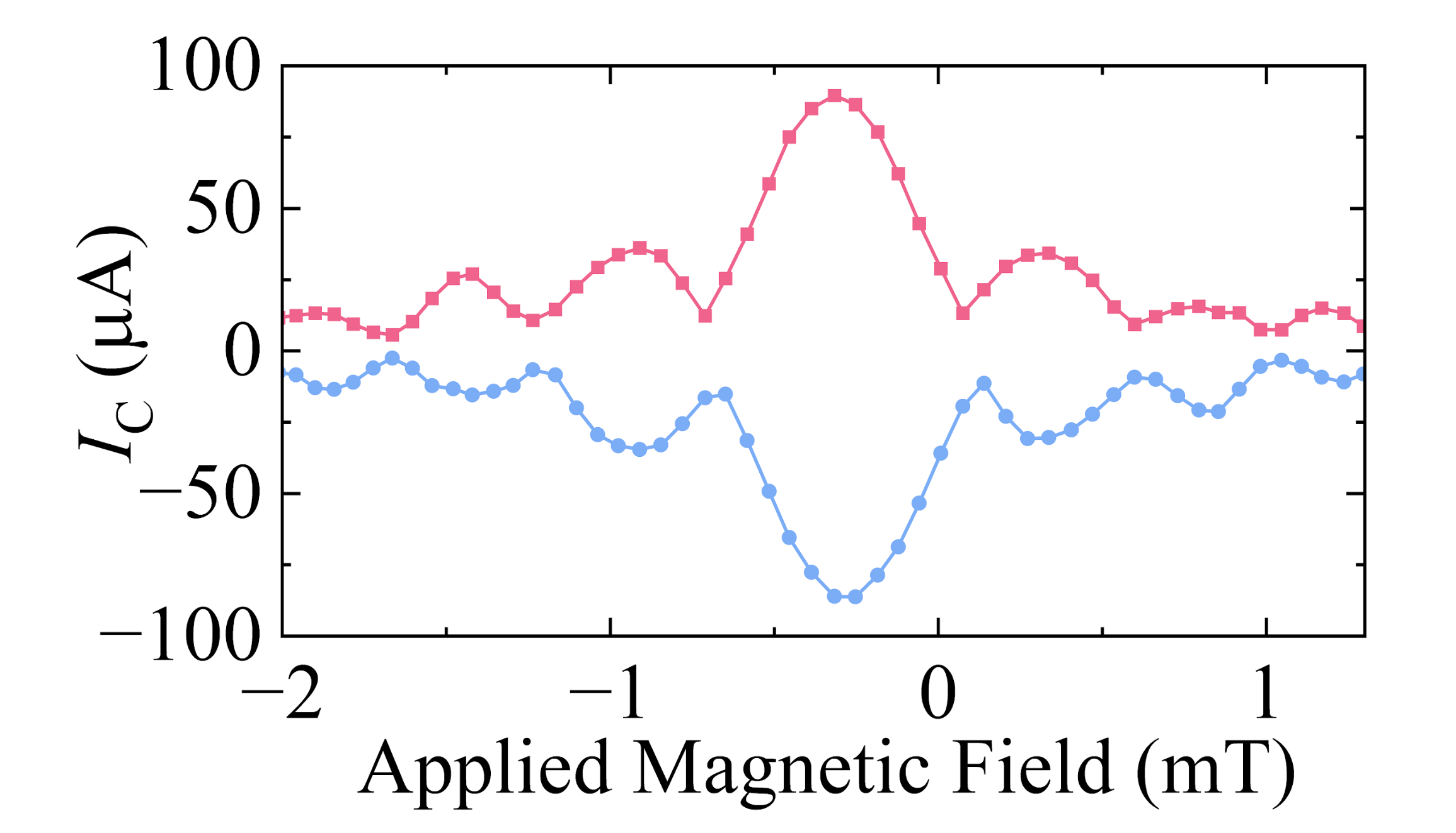}
\caption{Critical current versus applied magnetic field for the \SI{500}{ions/nm} Josephson junction at \SI{35}{K}, for positive (red squares) and negative (blue circles) bias, measured with the mu-metal shield removed. The field is applied perpendicular to the film. The pattern is centered near \SI{-0.3}{mT} because of the Earth's field and the remanent field of the PPMS magnet.}
\label{fig:IcB}
\end{figure}

Magnetic field modulation of the critical current is the direct test of the dc Josephson effect, and it provides an independent check of the $I_{\mathrm{C}}$ obtained from the $\mathit{IV}$ characteristics. Figure~\ref{fig:IcB} shows the critical current versus applied field at \SI{35}{K}, measured with the mu-metal shield removed. $I_{\mathrm{C}}$ oscillates with field for both current polarities, with a central maximum and a series of side lobes. The clear maxima and minima confirm coherent supercurrent transport across the barrier. The junction is a planar in-line junction in a thin film with strong flux focusing from the electrodes. Its pattern is therefore not expected to follow the $\sin(x)/x$ form of a sandwich junction with uniform current density, and we do not fit it to that form. The center of the pattern is shifted from zero applied field by approximately \SI{-0.3}{mT}. With the shield removed the sample sits in the Earth's field, and the PPMS magnet retains a remanent field after being energized. We attribute the offset to the background field at the sample and not to an asymmetry of the junction. The critical current at the central maximum is 90 \si{\micro\ampere} for both polarities. This is about 40 percent higher than the approximately \SI{65}{\micro\ampere} obtained from the shielded $\mathit{IV}$ measurement at the same temperature in Fig.~\ref{fig:IVT}(b). The two measurements were performed in separate runs. A plausible explanation is that the maximum of the modulation pattern marks the true zero of field at the junction, whereas the shielded $\mathit{IV}$ characteristics were measured in whatever residual field remained inside the shield, which places them slightly off the maximum. The values of $I_{\mathrm{C}}R$ in Fig.~\ref{fig:IVT}(c) are therefore, if anything, conservative.

The modulation pattern also bounds the excess current. Excess current from quasiparticle or Andreev processes does not modulate with field, so it would appear as a field independent floor beneath the interference pattern. In Fig.~\ref{fig:IcB} the critical current at the minima falls to 10 \si{\micro\ampere}. This is the measurement floor set by the 5 \si{\micro\volt} voltage criterion and the noise of the measurement, and a small residual at the minima is always present for this reason. Compared with the full peak-to-peak modulation of \SI{180}{\micro\ampere} between the positive and negative bias maxima, the excess current is therefore no larger than about 5 percent. This agrees with the extrapolation of the high-bias $\mathit{IV}$ branch. Together, the two measurements show that the transport is dominated by the phase coherent Josephson current.

\subsection{Microwave Response}

\begin{figure}[!t]
\centering
\includegraphics[width=0.46\textwidth]{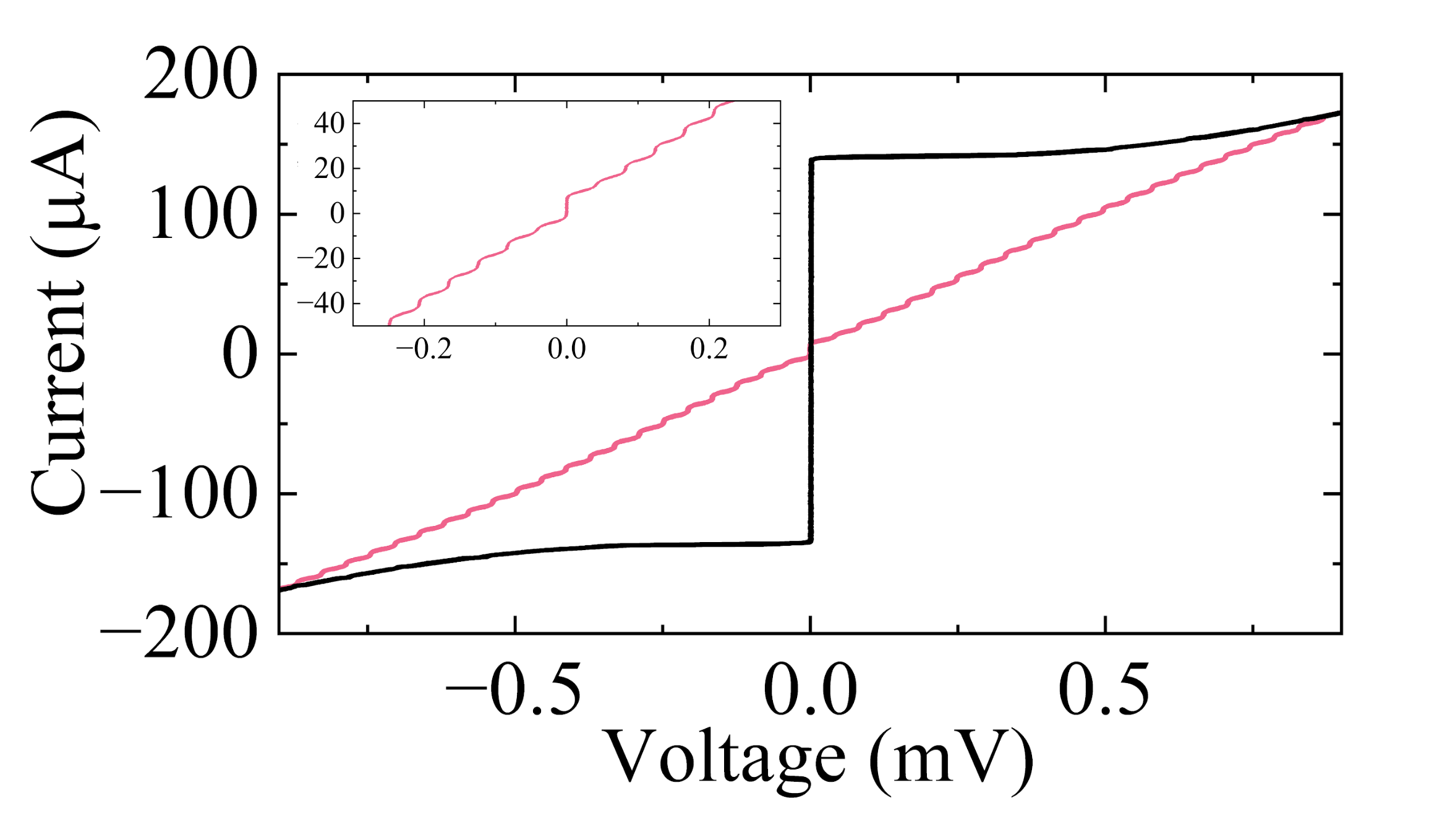}
\caption{Current-voltage characteristics of the \SI{500}{ions/nm} Josephson junction at \SI{20}{K} with (red) and without (black) \SI{20.5}{GHz} microwave irradiation, measured with the mu-metal shield removed. Inset: expanded view of the irradiated curve showing Shapiro steps spaced by $hf/2e = \SI{42.4}{\micro\volt}$.}
\label{fig:RF}
\end{figure}

Figure~\ref{fig:RF} shows the $\mathit{IV}$ characteristic under microwave irradiation. With \SI{20.5}{GHz} applied at \SI{12}{dBm} and \SI{20}{K}, the zero voltage critical current is fully suppressed, and the $\mathit{IV}$ characteristic develops a staircase of constant voltage Shapiro steps at integer multiples of $hf/2e = \SI{42.4}{\micro\volt}$, visible in the inset. Steps appear only at integer multiples of this voltage, consistent with a sinusoidal current phase relation. The steps are narrow relative to the full $\mathit{IV}$ characteristic because the drive frequency is far below the Josephson frequency of this junction. For a characteristic voltage of several hundred microvolts, the Josephson frequency $f_c = 2eI_{\mathrm{C}}R/h$ exceeds \SI{100}{GHz}, so the normalized drive frequency $f/f_c$ is much less than one. In this limit, the junction follows the drive adiabatically and the step spacing $hf/2e$ is small compared with $I_{\mathrm{C}}R$. The step widths oscillate with microwave amplitude, and at the amplitude shown the zero voltage step is near a null. Wider steps and stronger phase locking would be obtained at drive frequencies approaching $f_c$, which is the regime relevant for high frequency applications.\cite{Parachikunnumal.2026}

\subsection{Comparison With Previous Work}

Focused helium ion beam fabrication of YBCO Josephson junctions was first demonstrated in Ref.~\cite{Cybart.2015}, and it has since been studied across a range of device geometries, film thicknesses, and irradiation conditions.\cite{Propper.2026,Muller.2019,Cho.2018,Chen.2022} The characteristic voltages measured here between \SI{1.6}{K} and \SI{50}{K} are comparable to those reported for He FIB YBCO junctions in the literature, and they are consistently higher than those of the junctions we fabricated previously over the same temperature range. We remark that the definitions of $I_{\mathrm{C}}$ and $R$ used to form the product vary from report to report. The values here use the $I_{\mathrm{C}}$ from the shielded $\mathit{IV}$ characteristics with a 5 \si{\micro\volt} criterion and the high-bias slope $R$ defined in Section~\ref{sec:intro} and fitted over the 1 to 2 \si{\milli\volt} window given in Section~\ref{sec:iv}.

\section{Conclusion}

In conclusion, we have fabricated in-plane He FIB Josephson junctions in YBCO thin films with a modified process in which the exposed YBCO was cleaned after the final lithography step, only the cleanest samples were used, the beam focus and astigmatism were verified in every write field, and the samples were stored with desiccant. The magnetic field modulation of the critical current shows Josephson behavior with minima at the measurement floor, and the high-bias branch of the $\mathit{IV}$ characteristic extrapolates through the origin. Both results show that the excess current is negligible. Shapiro steps at $hf/2e$ under \SI{20.5}{GHz} irradiation confirm the ac Josephson effect. The characteristic voltage reaches 929 \si{\micro\volt} at \SI{1.6}{K}, falls within the range reported for similar devices between \SI{1.6}{K} and \SI{50}{K}, and is larger than that of our previous junctions. The temperature dependence of $I_{\mathrm{C}}$ is proximity-like near the junction transition and tunnel-like at low temperature. We attribute this to the barrier sitting at the metal insulator transition, where hopping conduction through localized states is active at finite temperature and freezes out on cooling. Which of the process changes is responsible for the improvement in $I_{\mathrm{C}}R$ remains to be determined and merits additional study. Larger characteristic voltages at these dimensions are an encouraging result for RSFQ and AQFP circuits and for SQUIDs operating at elevated temperature.

\section*{Acknowledgment}

The authors are grateful to the late Robert C. Dynes for the interpretation of the barrier transport presented in Section~\ref{sec:iv}.

\ifCLASSOPTIONcaptionsoff
  \newpage
\fi

\bibliographystyle{IEEEtran}
\bibliography{biblio}

\end{document}